\documentstyle[aps,12pt,amssymb]{revtex} \tighten 
\begin{document}

\newcommand{\p}{\partial} \newcommand{\g}{\gamma}
\newcommand{\Ld}{\Lambda} \newcommand{\ld}{\lambda}
\newcommand{\z}{\zeta} \newcommand{\be}{\begin{equation}}
\newcommand{\ee}{\end{equation}} \newcommand{\bea}{\begin{eqnarray}}
\newcommand{\eea}{\end{eqnarray}}

\title{Linearized Einstein theory via null surfaces}

\author{Simonetta Frittelli$^a$ \and Carlos N. Kozameh$^b$ \and Ezra T.
Newman$^a$} \address{${}^a$University of Pittsburgh, 100 Allen Hall,
Pittsburgh, PA 15260\\
	${}^b$FaMAF, Universidad Nacional de C\'ordoba, 5000
	C\'ordoba, Argentina} \maketitle \begin{center} DRAFT
\end{center}

\begin{abstract}

Recently there has been developed a reformulation of General Relativity
- referred to as {\it the null surface version of GR} - where instead
of the metric field as the basic variable of the theory, families of
three-surfaces in a four-manifold become basic. From these surfaces
themselves, a conformal metric, conformal to an Einstein metric, can be
constructed. A choice of conformal factor turns them into Einstein
metrics.  The surfaces are then automatically characteristic surfaces
of this metric.

In the present paper we explore the linearization of this {\it null
surface theory} and compare it with the standard linear GR. This allows
a better understanding of many of the subtle mathematical issues and
sheds light on some of the obscure points of the null surface theory.
It furthermore permits a very simple solution generating scheme for the
linear theory and the beginning of a perturbation scheme for the full
theory.

\end{abstract}

\section{Introduction}

It is our intention to introduce a new point of view to the Einstein
theory of gravity. We switch our attention from the metric tensor,
conventionally viewed as the fundamental field of the theory, to a
different set of quantities that are to be more basic than the metric
itself, from which the metric can  be extracted as a derived concept.
The quantities that are our candidates to serve this purpose are {\it
null (hyper)surfaces} of the spacetime. Null surfaces encode the
information of a Lorentzian metric up to a conformal factor and the
surfaces are simply sets of points, that do not carry the elaborate
behavior of tensor fields on a manifold, therefore being (perhaps)
suitable for a {\it more fundamental} description of the spacetime from
which additional structure can be derived, if desired. Our idea is to
obtain a description of gravity in which we are able to understand
Einstein spacetimes in terms of null surfaces. In this sense, we depart
from the more established view of thinking of GR as a theory for the
metric as a field tensor $g_{ab}$ on a manifold. We refer to this new
approach as a ``surface theory" of General Relativity.

The essence of the null surface approach to gravity (first introduced
in 1983~\cite{kn83}) is the use of a sphere's worth of null coordinate
systems which allows (up to a conformal factor) for all the components
of the metric tensor to be obtained by only differential and algebraic
manipulations on a \underline {single function}. The idea is to
introduce three parameters to label null surfaces in the manifold,
i.e.  a two parameter family of foliations.  Since every one-parameter
family of surfaces can be used to define a coordinate system (e.g. the
past light cones emanating from a world line), then we are actually
enlarging our set of variables by adding two more variables to the
coordinates. These two additional variables can be taken as coordinates
on the unit sphere.

More specifically, we begin with a Lorentzian manifold $M$ with a
metric $g_{ab}$, a given three parameter family of null hypersurfaces
on the manifold, namely a function of the form $F(x^a,u,\z,\bar\z)=0$,
the three parameters being $u$, $\z$ and $\bar\z$, or alternatively by
$u=Z(x^a,\z,\bar\z)$ with $g^{ab}Z,_aZ,_b=0$.  Then every set of null
surfaces with fixed value of $\z$ and $\bar\z$ (complex stereographic
coordinates on $S^2$), a foliation, can be chosen as the level surfaces
of a null coordinate $u$. If we apply an $\eth$ operation (essentially
$\frac{\p}{\p\z}$) to the function $Z$, we obtain another function of
$x^a$, $\z$ and $\bar\z$ which for fixed $\z$ and $\bar\z$ is in
general independent of the previous one. We define this as a second
coordinate $\omega$.  In the same fashion, the remaining two
coordinates of a new coordinate system for fixed $\z$ and $\bar\z$ can
be defined by applying $\bar\eth$ and $\eth\bar\eth$ to the original
function $Z$, and will be denoted $\bar\omega$ and $R$ respectively.
What we obtain is a new coordinate system for every value of the
parameters $\z$ and $\bar\z$, i.e., a sphere's worth of coordinate
systems.  The structure we have is that of the bundle of null
directions over $M$, with each fiber an $S^2$ and a coordinate system
associated with each fiber point. There is a major advantage to having
this ``bundle" of coordinate systems:  one can differentiate the
components of any tensor with respect to the parameters, i.e.
vertically,  and obtain relationships among the components themselves.
In other words, the components of a tensor will not be independent of
each other, and it will be feasible to obtain most of them by knowledge
of only one for all values of $(\z,\bar\z)$.  This and the fact that
one of the coordinates is null provide the basis for describing the
metric tensor completely in terms of the single function
$Z(x^a,\z,\bar\z)$, which becomes a basic variable of null surface
theory.

The dynamics of the function $Z$ has been a troublesome issue ever
since its appearance. In principle, all 10 of Einstein equations are
encoded in only one equation~\cite{b-kn84} for the conformal factor
$\Omega$ and $Z(x^a,\z,\bar\z)$ by virtue of the sphere's worth of
coordinate systems. There are, in addition, several ``auxiliary"
conditions that guarantee that the metrics, eventually extracted from
$Z(x^a,\z,\bar\z)$, are all diffeomorphic to a metric $g_{ab}(x^c)$.
Considerable success was originally attained by introducing the
holonomy operator around closed loops as an auxiliary
variable~\cite{i93}, though the resulting theory was quite
complicated.  Recently a great simplification was achieved where the
full GR was expressed as equations for $Z$ and $\Omega$. There however
still remained certain difficult conceptual issues, which led us to
study the linearized version developed below.  This gives a new insight
into the structure of the full equations.

We first review, in Sec.II, the conventional approach (conventional as
opposed to the null surface approach) to linearized gravity, showing
the Einstein equations in terms of the first order correction to an
assumed underlying flat space. We make a distinction between trace and
trace free part of the linearized metric and give the equations in
terms of these two quantities instead of the full first order tensor.
We then show how a choice of gauge, in the conventional sense, can lead
to a new coordinate system that is similar to the one described above,
except for the critical $(\z,\bar\z)$ dependence.  We believe that this
comparison of the conventional view with the null surface view
clarifies many of the issues of null surface theory in first order.

A complete discussion of first order null surface theory for gravity
with sources is found in sections III and IV. The attention is focused
on the role of sources and a possible cosmological constant on the two
relevant quantities. We find that all the components of the metric can
be expressed algebraically in terms of derivatives of a single
spin-weight-2 function denoted $\Ld$, which is defined by
$\Ld\equiv\eth^2Z$, and the trace of the metric $\nu$ which to first
order is equivalent to a small correction to the conformal factor
$\Omega$. It is remarkable that the metric does not depend directly on
$Z$ but only through $\Ld$, which means that the metric does not know
of the hypersurfaces themselves, but of a related quantity $\Ld$ whose
geometrical meaning will be discussed elsewhere.

\section{The linearized field equations}

We begin with a spacetime and a metric $g_{ab}$ that differs from a
flat metric by a small term $\g_{ab}$ with $g_{ab}=\eta_{ab}-\g_{ab}$.
In a perturbation expansion to first order, the Einstein equations for
$g_{ab}$ yield a system of linear second order equations for $\g_{ab}$
on the flat background $\eta_{ab}$. In particular, the linearized Ricci
tensor becomes~\cite{w84}:

\be R_{ab} = -\p^c\p_{(b}\g_{a)c} + \frac12 \Box\g_{ab} + \frac12
\p_a\p_b\g \ee

\be R = -\p^c\p^d\g_{cd} + \Box\g \ee where $\p_a$ is the flat
derivative, $\Box \equiv \p^c\p_c$, $R$ and $\g$ are the (linearized)
traces of the Ricci tensor and the first order correction to the flat
metric. Indices are raised and lowered with $\eta_{ab}$.  We decompose
the first order metric into its trace-free part and its trace:
$\g_{ab}=\frac14\g\eta_{ab}+q_{ab}$ and redefine the trace $\g$ in
terms of a scalar $\nu$ in the following way: $\g=8\nu$.  The Ricci
tensor, in terms of the redefined ``trace" $\nu$ and the trace-free
metric $q_{ab}$, becomes \be R_{ab} = 2\p_a\p_b\nu + \eta_{ab}\Box \nu
		      - \p^c\p_{(b}q_{a)c} + \frac12\Box q_{ab}
\ee and the Ricci scalar becomes:

\be R =  6\Box \nu -\p^c\p^d q_{cd}.  \ee

We assume that the source of the Einstein equations
$R_{ab}-\frac12g_{ab}R=2T_{ab}$ is a first order stress-energy tensor
that is conserved, and  does not depend on the first order correction
to the metric, but on the flat metric at most. As defined here,
$T_{ab}$ absorbs the factor $\kappa /2$ that is usually explicit.
$T_{ab}$ can be written as $T_{ab} = t_{ab} + \frac14\eta_{ab}T$ with
$t_{ab}$ trace free and $T$ its trace. The conservation law
$\p^aT_{ab}=0$ for the stress energy tensor in terms of $t_{ab}$ and
$T$ takes the form:

\be
	\p_bT = -4\p^at_{ab}.           \label{T} \ee

{\it We will refer to $q_{ab}$ loosely as the metric and to $\nu$ as
the trace}, to avoid unnecessary complexity in the terminology which
may obscure the content of the work.

The complete linearized Einstein equations in terms of $\nu$ and
$q_{ab}$ read:

\be
 \p_a\p_b\nu - \eta_{ab}\Box \nu - \frac12\p^c\p_{(b} q_{a)c} + \frac14
\Box q_{ab} + \frac14 \eta_{ab} \p^c\p^d q_{cd} = T_{ab}.   \label{EM}
\ee

It is possible to choose coordinate systems so that the flat metric has
only the following nonvanishing components:  $\eta_{00}\!=\!2$,
$\eta_{01}\!=\!1$, and $\eta_{+-}\!=\!-1$. A way to do this is by
choosing a null tetrad $(l_a, n_a, m_a, \bar m_a)$ where $l_a$ and
$n_a$ are real, $m_a = a_a + ib_a$ with $a_a$ and $b_a$ real spacelike
vectors, and all the scalar products being zero except $l^an_a=-m^a\bar
m_a=1$. Then we would take $l_a$, $m_a$, $\bar m_a$ and the combination
$n_a\!-l_a$ as our coordinate basis vectors. The inverse $\eta^{ab}$
has the following non-vanishing elements:  $\eta^{11}\!=\!-2$,
$\eta^{01}\!=\!1$ and $\eta^{+-}\!=\!-1$.  We will understand the
equations (\ref{EM}) as equations for the two quantities $\nu$ and
$q_{ab}$ in these particular flat coordinates, so that both of them are
functions of the coordinates $(x^0,x^1,x^+,x^-)$ and partial
derivatives are taken with respect to these coordinates. An example of
these coordinates would be $x^0=z-t$, $x^1=z$, $x^+=x+iy$ and
$x^-=x-iy$.

The linearized equations (\ref{EM}) admit gauge freedom for both the
metric and the trace in the following form:

\be q'_{ab} = q_{ab} + \p_a\xi_b + \p_b\xi_a
				 - \frac12\eta_{ab}\p^c\xi_c
							\label{gaugeq}
\ee

\be \nu' = \nu - \frac14 \p^a\xi_a.                     \label{gaugenu}
\ee where $\xi_a$ is a free vector field.

Using this gauge freedom we can make $q'_{1b}=0$ for $b=1,+,-$ and
$q'_{01}\!+q'_{+-}=0$.  Notice that since $q'_{ab}$ is trace free then
$0=\eta^{ab}q'_{ab}=2(q'_{01}\!-q'_{11}\!-q'_{+-})$ implies
$q'_{+-}=q'_{01}=0$ as well. There are no restrictions on the trace
$\nu$. These conditions on the gauge can be expressed as:

\[ \p_1\xi_b + \p_b\xi_1 - \frac12\eta_{1b}\p^c\xi_c + q_{1b} = 0, \]
which can be solved explicitly with a given $q_{1b}$.  There is still
the remaining gauge freedom of the homogeneous system:

\[ \p_1\xi_b + \p_b\xi_1 - \frac12\eta_{1b}\p^c\xi_c = 0.  \] The
solutions to this system leave four unspecified functions of the
coordinates $x^0$, $x^+$ and $x^-$, while the dependence on $x^1$ is at
most quadratic. The corresponding divergence $\p^c\xi_c$ is linear in
$x^1$ with two free functions of the remaining coordinates as the
coefficients. We will return to the issue of the gauge freedom in
Sec.IV when we introduce a special family of coordinate systems and
restrict to the transformations within the family.

Summarizing, the linearized metric has the form:  \[ g_{ab}=
\Omega^{-2} (\eta_{ab} - q_{ab}) =  (1-2\nu)\left\{ \left(
\begin{array}{cccc} 2 & 1 & 0 & 0 \\ 1 & 0 & 0 & 0 \\ 0 & 0 & 0 & -1 \\
0 & 0 & -1 & 0 \end{array} \right) - \left( \begin{array}{cccc} q_{00}
& 0 & q_{0+} & q_{0-} \\ 0 & 0 & 0 & 0 \\ q_{0+} & 0 & q_{++} & 0 \\
q_{0-} & 0 & 0 & q_{--} \end{array} \right) \right\} \] with inverse:
\[ g^{ab}= \Omega^2 (\eta^{ab} + q^{ab}) =  (1+2\nu)\left\{ \left(
\begin{array}{cccc} 0 & 1 & 0 & 0 \\ 1 & -2 & 0 & 0 \\ 0 & 0 & 0 & -1
\\ 0 & 0 & -1 & 0 \end{array} \right) + \left( \begin{array}{cccc} 0 &
0 & 0 & 0 \\ 0 & q^{11} & q^{1+} & q^{1-} \\ 0 & q^{1+} & q^{++} & 0 \\
0 & q^{1-} & 0 & q^{--} \end{array} \right) \right\} \]

If we now look at the $a=1$, $b=1$ component of the Einstein equations
(\ref{EM}) we obtain a single equation that relates the sources to only
the trace of the first order metric, since all the terms that involve
the trace free part $q_{ab}$ are identically zero by virtue of our
choice of gauge. The equation will be:  \be
	\p_1^2 \nu = t_{11}.            \label{nu11} \ee

This equation tells us that the source directly drives $\nu$, i.e.
$\nu$ is given in terms of a double integral of the source $t_{11}$
plus any solution of the homogeneous equation. Note that the trace free
component $t_{11}$ is equal to the full $T_{11}$.

The homogeneous solutions to (\ref{nu11}) are all functions that are
linear in the coordinate $x^1$, with two free functions of the
remaining coordinates as the coefficients, i.e. $\nu_H = \alpha + \beta
x^1$. We will return to this later.

The motivation for the discussion up to now has been to establish a
basis for the understanding of {\bf first order null surface theory}
from a more conventional point of view. We will see that, in a sense to
be discussed, Eq.(\ref{nu11}) {\bf is the single equation that encodes
the dynamics of first order Einstein fields from the null surface point
of view}. The new ingredient in null surface theory that explains why
the remaining nine equations do not give any other relevant information
is that it makes exhaustive use of the introduction, \underline{not} of
a single coordinate system with these properties, but a smooth
two-parameter family of them. We will devote the following sections to
a rather complete discussion of the ideas of null surface theory to
first order.

\section{A sphere's worth of coordinate systems}

We begin with a manifold $M$ with a Lorentzian metric $g_{ab}$ (we
might also want a boundary to $M$ if we wish to discuss asymptotically
flat spaces). A choice of coordinates based on null surfaces is made in
the following fashion. We choose a family of null surfaces that are
labeled by points of a three parameter space:  $R\times S^2$, with a
real parameter $u$ on $R$, and two stereographic coordinates
$(\z,\bar\z)$ on $S^2$. This is expressed by $F(x^a,u,\z,\bar\z)=0$, or
preferably by $u=Z(x^a,\z,\bar\z)$. An immediate meaning to the
function $Z$ from this point of view is that for every fixed value of
the parameters $(\z,\bar\z)$ the level surfaces of $u=Z$ are null
surfaces i.e.  $g^{ab}\p_aZ\p_bZ=0$, labeled by the parameter $u$.

For the construction of our coordinate system we choose four functions
of the points $x^a$ that are simply $Z$, $\eth Z$, $\bar\eth Z$, and
$\bar\eth\eth Z$ at an arbitrary but fixed value of the parameters
$(\z,\bar\z)$.  We choose to work with the $\eth$ and $\bar\eth$
operators instead of ordinary partial derivatives because of their
covariant nature~\footnote{We assume familiarity with the $\eth$ and
$\bar\eth $ operators. See reference ~\cite{np66}.}. We thus define an
$S^2$'s set of coordinate systems for the spacetime in the following
way:  \bea
	 u &  = &  Z(x^a,\z,\bar\z)                     \nonumber\\
    \omega &  = &  \eth Z(x^a,\z,\bar\z)                \nonumber\\
\bar\omega &  = &  \bar\eth Z(x^a,\z,\bar\z)            \nonumber\\
	 R &  = &  \bar\eth\eth Z(x^a,\z,\bar\z)        \label{coord}
\eea where the parameters $(\z,\bar\z)$ have fixed values (after
differentiation). As a matter of notation we refer to them as $\theta^i
= (u,R,\omega,\bar\omega)$, with $i = 0,1,+,-$. We thus have $\theta^i
= \theta^i(x^a,\z\bar\z)$.

[As an example of this construction in an asymptotically flat spacetime
we can think of the interior spacetime points as being $x^a$, while the
boundary is a three-dimensional space (the null surface $\cal J^+$)
which can be given the coordinates $u$, $\z$ and $\bar\z$. From a fixed
value of $(u,\z,\bar\z)$, the past light cone is the set of points
$x^a$ that satisfy $Z(x^a,\z,\bar\z)=u$, thus $u$ labels past light
cones from fixed null generators of $\cal J^+$. Early history of the
null surface theory gave $Z$ the name {\it light cone cut
function}~\cite{kn83}; we will sometimes refer to $Z$ as the cut
function. On the given past light cone, the values of
$(\omega,\bar\omega)$ label the null generators, thus a particular
value of $(u,\omega,\bar\omega)$ is a null geodesic emanating from the
point $(u,\z,\bar\z)$ at $\cal J^+$. The last coordinate $R$ is a
parameter along the null geodesics (in general not affine).]

Two different values of $(\z,\bar\z)$ give two different coordinate
systems. However, whatever coordinate conditions are implied by
(\ref{coord}) will be true for all values of $(\z,\bar\z)$. Since, by
assumption, the dependence on $(\z,\bar\z)$ is smooth, one can go back
and forth from one coordinate system to a neighboring one by
differentiating with respect to the parameters $(\z,\bar\z)$, or
equivalently, by applying $\eth$ and $\bar\eth$ operations.

 From (\ref{coord}) we obtain a sphere's worth of coordinate tetrads.
The gradients $\theta^i_{,a}$ give us a coordinate set of covectors:

\bea
 \theta^0_{,a} &  = &  Z_{,a}
 \nonumber\\ \theta^+_{,a} &  = &  \eth
 Z_{,a}                              \nonumber\\ \theta^-_{,a} &  = &
 \bar\eth Z_{,a}                          \nonumber\\ \theta^1_{,a} &
 = &  \bar\eth\eth Z_{,a}.  \eea

The coordinate tetrad vectors $\theta_i^a$ dual to $\theta^i_{,a}$ are
defined by $\theta^i_{,a}\theta_j^a = \delta_i^j$.

We will be concerned with the linearized version of the Einstein
equations, in which all terms are flat second derivatives of the first
order correction to the flat metric. The function $Z$ to first order
consists of a zeroth order $Z_M$ and a first order term $z$, i.e.  \be
Z(x^a,\z,\bar\z) = Z_M(x^a,\z,\bar\z) + z(x^a,\z,\bar\z) \ee where
$Z_M$ is supposed to be known and $z$ becomes the first order variable.
The role of $Z_M$ is to fix the zeroth order flat background:  the flat
intrinsic coordinates and tetrad. It will not be necessary to keep
first order terms in the tetrad, since, in what follows, they will
always appear contracted with tensors that are already first order (see
remark below). We will work with the flat tetrad obtained from the
function $Z_M$ for flat space defined by a sphere's worth of null
planes, which is easily seen to be~\cite{kn83}:

\be Z_M = x^al_a(\z,\bar\z) \ee where $x^a$ are standard Minkowskian
coordinates and \be l^a(\z,\bar\z) = \frac{1}{\sqrt2(1+\z\bar\z)}
\biggl((1+\z\bar\z),\;\z+\bar\z,\;i(\bar\z-\z),\;-1+\z\bar\z\biggr)
				\label{l}
\ee is a null vector for all $(\z,\bar\z)$ with respect to the flat
$\eta_{ab}$.  The index is raised with $\eta_{ab}$ with signature
$(+---)$. It is customary to define a null tetrad out of this function
where the remaining vectors are:  \bea
	 m_a & = & \eth l_a                     \nonumber\\ \bar m_a &
    = & \bar\eth l_a                 \nonumber\\
	 n_a & = & \bar\eth\eth l_a+l_a         \label{ethm}
\eea so that $n^al_a\!=\!1$, $m^a\bar m_a\!=\!-1$ and the remaining
scalar products are zero. Following our approach, we define
$\theta^i=(u,R, \omega,\bar\omega)$ as in (\ref{coord}). Then:  \bea
	u & = & x^a l_a(\z,\bar\z)              \nonumber\\ \omega & =
	& x^am_a(\z,\bar\z)                  \nonumber\\ \bar\omega & =
	& x^a\bar m_a(\z,\bar\z)         \nonumber\\ R & = & x^a
	(n_a(\z,\bar\z)-l_a(\z,\bar\z)).  \eea Our tetrad is then:

\be
   \theta^i_{,a} = \bigl( l_a, \;n_a-l_a,\; m_a,\; \bar m_a\; \bigr)
						\label{tetrad}
\ee

and dual:  \be
   \theta_i^a = \bigl( n^a+l^a,\; l^a,\; -\bar m^a,\; - m^a \;\bigr)
						\label{dual}
\ee and the partial derivatives with respect to the coordinates
$\theta^i$ are $\p_i = \theta_i^a\p_a$.

The flat metric $\eta_{ij}$ in these coordinates has three
non-vanishing elements:  $\eta_{00}\!=\!2$, $\eta_{01}\!=\!1$, and
$\eta_{+-}\!=\!-1$ with the inverse $\eta^{ij}$ having the elements:
$\eta^{11}\!=\!-2$, $\eta^{01}\!=\!1$ and $\eta^{+-}\!=\!-1$
non-vanishing, i.e., is of the form we used in Sec.II.

We end this section by pointing out that this $(\z,\bar\z)$-dependent
set of coordinate systems has the properties (i.e. satisfies the same
set of coordinate conditions as in Sec.II.) required to obtain the
equation for the trace $\nu$ of the first order metric from the
Einstein system of equations, namely Eq.(\ref{nu11}),
$\p_1^2\nu=t_{11}$, for each value of $(\z,\bar\z)$.  The advantage now
is that we have a smooth two-parameter family of such coordinate
systems, so we are allowed to perform differentiations with respect to
$(\z,\bar\z)$ without changing the coordinate conditions. These remarks
will be amplified upon in the next section.

\section{First order null surface theory}

\subsection{The null theory variables $\nu$ and $\Ld$ and kinematic
equations}

The components of any tensor referred to the new coordinates are
obtained by contraction of the old tensor with the basis or dual
tetrad. In particular, the metric components with respect to the new
coordinates denoted by $g_{ij}$ are given by
$g_{ij}=g_{ab}\theta_i^a\theta_j^b$, with inverse
$g^{ij}=g^{ab}\theta^i_{,a}\theta^j_{,b}$.

There is a distinction between $ij$ and $ab$ indices.  Quantities like
$g_{ab}$ or $T_{ab}$ are ``local" in the sense that they depend only on
the local coordinates of the spacetime $x^a$.  However $g_{ij}$ or
$T_{ij}$ depend on $\theta^i$ and $(\z,\bar\z)$ as well.  Similarly,
the partial derivatives $\p_a$ commute with $\eth$ and $\bar\eth$,
whereas $\p_i$ do not in general.  The important commutation relations
of $\eth$ and $\bar\eth$ with $\p_i$ are given later.

In these coordinates~\cite{kn83}, all the components of the metric can
be expressed in terms of two functions. One of these two functions is
defined as follows:  \be \Ld \equiv \eth^2 Z.  \ee All the components
of the metric get expressed as $g^{01}$ times a simple algebraic
function of $\Ld$ and its derivatives. Therefore $g^{01}$ can be taken
as a {\em conformal factor\/} $\Omega^2=g^{01}
=g^{ab}\theta^0_{,a}\theta^1_{,b}=g^{ab}Z,_a\eth\bar\eth Z,_b$. Thus
the metric is conformally a function of $\Ld$, the conformal factor
being $g^{01}$. This follows essentially from the fact that $g^{00} =
g^{ab}Z,_aZ,_b$ is zero by assumption, since $u=Z=constant$ is a null
surface, and the remainder of the components can be computed by
applying $\eth$ and $\bar\eth$ to $g^{00}=0$ an appropriate number of
times, {\it using the non-trivial fact that $\eth g^{ab}=\bar\eth
g^{ab}=0$}. For example, from \be
 g^{00}=g^{ab}Z,_aZ,_b=0 \ee and \be g^{0+}=g^{ab}Z,_a\eth Z,_b, \ee by
applying $\eth$ to $g^{00}=0$ we obtain \be \eth g^{00}=2g^{0+} \ee
 and therefore $g^{0+}=0$.  Applying $\eth\bar\eth (g^{ab}Z,_aZ,_b)=0$
we get $\eth\bar\eth g^{00}=2g^{01}+2g^{+-}=0$ and therefore \be
g^{+-}=-g^{01}.  \ee Taking $\eth^2\! g^{00}=0$ gives the expression of
$g^{--}$ in terms of $g^{01}$ and $\Ld$.  Taking $\eth\bar\eth^2$ of
$g^{00}=0$ gives $g^{1+}$.  Finally, by taking $\eth^2\bar\eth^2$ of
$g^{00}$ we obtain $g^{11}$.  The complex conjugates of all these
operations yield the remaining components of the metric.

The linearization can be carried out by noticing that $Z_M = x^al_a$
satisfies \[ \eth^2 Z_M = \bar\eth^2 Z_M = 0.  \] Thus the zeroth order
does not contribute to $\Ld$ and therefore $\Ld$ is a first order
quantity.  By dividing the $g^{ij}$ by $g^{01}$ and linearizing in
$\Ld$ we have ${g^{ij}}/{g^{01}} = \eta^{ij} + f^{ij}$ where $f^{ij}$
is first order and  has the following five non-vanishing elements:

\bea f^{11} & = & -\frac12\bar\eth^2\Ld_1 + \bar\eth\Ld_+ =
		 -\frac12\eth^2\bar\Ld_1+ \eth\bar\Ld_-
		 \nonumber\\
f^{1+} & = & -\frac12 \bar\eth\Ld_1
\nonumber\\ f^{1-} & = & -\frac12
\eth\bar\Ld_1                             \nonumber\\ f^{++} & = &
-\Ld_1                                             \nonumber\\ f^{--} &
= & -\bar\Ld_1 \eea

The non-vanishing $f_{ij}$ are the following:  \bea f_{00} & = &
-\frac12\bar\eth^2\Ld_1 + \bar\eth\Ld_+ =
		 -\frac12\eth^2\bar\Ld_1+ \eth\bar\Ld_-
		 \nonumber\\
f_{0+} & = & \frac12 \eth\bar\Ld_1
\nonumber\\ f_{0-} & = & \frac12
\bar\eth\Ld_1                              \nonumber\\ f_{++} & = &
-\bar\Ld_1                                         \nonumber\\ f_{--} &
= & -\Ld_1                                     \label{metric} \eea
where we have adopted the notation $\p_i\Ld\equiv\Ld_i$. Note that the
first order trace-free metric is given in terms of $\Ld$ and its
derivatives.

In deriving the components of the metric from $g^{00}=0$ we used
repeatedly $\eth g^{ab}=\bar\eth g^{ab}=0$. This is the statement that
all the $(\z,\bar\z)$-dependent metrics obtained by
$g^{ij}=g^{ab}\theta^i_{,a}\theta^j_{,b}$ are in fact diffeomorphic to
$g^{ab}$, i.e., they are all equivalent. The conditions  $\eth
g^{ab}=\bar\eth g^{ab}=0$ constitute a set of 20 equations of which
some, when expressed in these coordinates, give the metric
(Eq.(\ref{metric})). Some are identities. The remaining equations,
given below, are referred to as the ``metricity conditions". We discuss
their meaning and significance in ~\cite{fkn95}, including an explicit
derivation of the metricity conditions from a slightly different point
of view.

Notice that $f=\eta^{ij}f_{ij} =0$ automatically. On the other hand,
the conformal factor $g^{01}=1+2\nu$ is equal to $\eta^{01}=1$ to
zeroth order plus a first order correction $2\nu$ which can be
identified with the trace of the full first order departure from flat
space, since
$g^{ij}=g^{01}(\eta^{ij}+f^{ij})=\eta^{ij}+2\nu\eta^{ij}+f^{ij}$ and
therefore the trace of the first order correction is indistinguishable
from the first order conformal factor. In fact, we will make no
distinction between the trace $\nu$ of section II and the first order
conformal factor of the current section.

We can continue to take $\eth$ and $\bar\eth$ of the components of the
metric that we have already obtained. The only new relations (the
metricity conditions) that we can obtain in this way~\cite{fkn95} are
the following:

(*) By writing $g^{01}=g^{ab}Z,_a\bar\eth\eth Z,_b$, applying $\eth$ to
$g^{01}$ and linearizing in $\Ld$ we obtain the relationship between
$\nu$ and $\Ld$:  \be
     2\eth\nu = \Ld_+ + \frac12 \bar\eth\Ld_1.          \label{ethg01}
\ee

(A) If we apply $\eth^3$ to $g^{00}=g^{ab}Z,_aZ,_b=0$ (the
 $\theta^+_{,a}\theta^+_{,b}$-component of $\eth g^{ab}=0$) and
 linearize in $\Ld$ we obtain the following relation for $\Ld$ only:

\be
       0 = \Ld_-  - \frac12 \eth\Ld_1.                   \label{A} \ee

(B) We can now take
$\bar\eth\eth^3g^{00}=\bar\eth\eth^3(g^{ab}Z_aZ_b)=0$ (the
$\theta^1_{,a}\theta^+_{,b}$-component of $\eth g^{ab}=0$). After
linearizing in $\Ld$ we obtain:  \be
	0 = \bar\eth(\eth\Ld_1 - 2\Ld_-).               \label{B} \ee

(C) Finally, by taking $\bar\eth^2\eth^3g^{00}=0$ (the
$\theta^1_{,a}\theta^1_{,b}$-component of $\eth g^{ab}$) we obtain
another linearized condition on $\Ld$:  \be
   0 = \eth^3\bar\Ld_1 -2 \eth\bar\eth\Ld_+ + 4\bar\eth\Ld_0
	-4\Ld_+ -2\bar\eth\Ld_1 \ee which, with the help of
(\ref{ethg01}) gives:

\be
   0 = \eth^3\bar\Ld_1 + \eth\bar\eth^2\Ld_1 + 4\bar\eth\Ld_0
	-8\eth\nu -4\eth^2\bar\eth\nu.          \label{C} \ee

These four first order relations (*), (A), (B) and (C), are of a
kinematical nature in the sense that they hold for any linearized
Lorentzian metric independent of the Einstein equations. They are very
important to the null surface approach, and as such they deserve a name
of their own: we refer to them as the {\it metricity conditions}.  If
the null surfaces $u=Z(x^a,\z,\bar\z)$ of a given metric are known and
$\Ld$ is computed, then the metricity conditions are identities. But if
the null surfaces are yet to be found, then the metricity conditions
become the equations that yield $\Ld$, from which the metric and, if
desired, the null surfaces as well, can be computed (the metricity
conditions are the requirement that the sphere's worth of metrics
$g^{ij}$ be really equivalent).  It should be noted that in first order
theory Eq.(B) follows from Eq.(A)~\footnote{B remains an identity even
in the full theory (see ~\cite{fkn95})}; this leaves only three
relevant metricity equations (*), (A) and (C).  They play an important
role later.

Before we further analyze these kinematic relations and properties of
the $f_{ij}$ we need the commutation relations between $\eth$ and the
$\p_i$. Since $ \eth\p_i F = \eth (\theta^a_i\p_aF) =
\eth\theta^a_i\p_aF + \theta^a_i \p_a\eth F$ and since $F$ will be
chosen as first order then:  \bea \p_0\eth & = &
\eth\p_0                                 \nonumber\\ \p_-\eth & = &
\eth\p_-                                 \nonumber\\ \p_+\eth & = &
\eth\p_+ + \p_0 - 2\p_1                  \nonumber\\ \p_1\eth & = &
\eth\p_1 + \p_-                          \label{cr} \eea with the
complex conjugate relations between $\bar\eth$ and the $\p_i$.  In
obtaining these relations we used the following:

\[ \eth\theta^0 = \eth Z \equiv \theta^+ \] \[ \eth\theta^+ = \eth^2Z =
0 \] \[ \eth\theta^- = \eth\bar\eth Z = \theta^1 \] \be \eth\theta^1 =
\eth^2\bar\eth Z = -2\theta^+.  \ee or alternatively relations
(\ref{ethm}), (\ref{tetrad}) and (\ref{dual}).

We mention here some very useful results that we will need later,
namely the divergence $\p^jf_{ij}$ and the double divergence
$\p^i\p^jf_{ij}$. With the use of the commutation relations  and the
kinematic relation (*) we can evaluate $\p^jf_{ij}$.  First we notice
that since $f_{1j} = 0$ then the contraction $\p^jf_{ij}$ has only
three components.  Also, since $\p^jf_{+j}$ is the complex conjugate of
$\p^jf_{-j}$ we only need to compute the components $i=0$ and $i=-$. We
display below the results and leave the detailed calculation to an
appendix:

\be \p^jf_{-j} = 2\p_1\eth\nu
\label{div-} \ee with the complex conjugate:  \be \p^jf_{+j} =
2\p_1\bar\eth\nu                                 \label{div+} \ee and
finally \be \p^jf_{0j} =  -2\p_1\bar\eth\eth\nu + 2\p_+\eth\nu
					+2\p_-\bar\eth\nu .
					\label{div0}
\ee

The double divergence $\p^i\p^jf_{ij}$ turns out to be simply:

\be
 \p^i\p^jf_{ij} = -2\p_1^2\bar\eth\eth\nu.
 \label{doublediv} \ee Note that although $f_{ij}$ are functions of
$\Ld$ the divergences are functions of only $\nu$.

\subsection{The dynamics of $\nu$ and $\Ld$}

So far the description of null surface theory has been completely
kinematical. Our variables ($\nu$ and $\Ld$) must satisfy the metricity
conditions (*), (A) and (C) to define for us a Lorentzian metric.  We
now address the problem of finding a metric that in addition satisfies
the Einstein equations. The Einstein equations are implemented in the
following way. We contract the linearized Einstein equations (\ref{EM})
with $l^al^b$ ($=\theta^a_1\theta^b_1$):  \be
 l^al^b\Bigl(\p_a\p_b\nu - \eta_{ab}\Box \nu - \frac12\p^c\p_{(b}
q_{a)c} + \frac14 \Box q_{ab} + \frac14 \eta_{ab} \p^c\p^d q_{cd}\Bigr)
= T_{ab}l^al^b.  \ee Because our gauge is the one described in section
II, and because $l^a$ is null, we immediately obtain:

\[
 l^al^b\p_a\p_b\nu =  T_{ab}l^al^b \] or \be
\p_1^2\nu(\theta^i,\z,\bar\z) =  t_{11}(\theta^i,\z,\bar\z). \label{e}
\ee This is {\it the} dynamical equation for $\Ld$ and $\nu$. Though it
resembles Eq.(\ref{nu11}), its meaning is different. Because of the
$(\z,\bar\z)$ dependence of $l^a$, it is not a single equation, but an
$S^2$'s worth of equations instead. It {\it contains} all the equations
that make up the trace-free part of the Einstein system.

[Another way to see that Eq.(\ref{e}) encodes the trace-free part of
the Einstein system is the following. We can apply $\eth$ and
$\bar\eth$ to Eq.(\ref{e}) an appropriate number of times so that we
obtain the remaining eight of the trace-free equations. It is clear
that we can do so because $t_{11}=T_{ab}l^al^b$, and thus an $\eth$
operation on $t_{11}$ goes through $T_{ab}$ and is applied to $l^a$
directly, giving a different tetrad vector ($m^a$), and therefore
giving a different component of $T_{ab}$. See the appendix for a more
complete discussion of this issue.]

It may appear that we are lacking a tenth equation, the trace of the
Einstein system. It is true that the trace is lost when we contract the
Einstein tensor with $l^al^b$. However, the information contained in
the trace of the Einstein system is redundant (see Eq.(\ref{T})). If we
put the Einstein equations in terms of the trace and trace-free part of
the Einstein tensor $G_{ab} = \underline{G}_{ab} + \frac14g_{ab}G$ they
become:

\[ \underline{G}_{ab} = t_{ab} \] and \[ G=T.  \] In addition, the
Bianchi identities $\bigtriangledown^aG_{ab}=0$ become:  \[
\bigtriangledown_bG=-4\bigtriangledown^a\underline{G}_{ab}.  \]
Therefore the trace of the Einstein tensor is determined (up to a
constant) by the Bianchi identities. However, the Bianchi identities
are of a kinematical nature, in that they are true for {\it any}
Lorentzian metric. This means that the trace of the Einstein system is
redundant when $\underline{G}_{ab}$ and $G$ are expressed as functions
of a metric. Now returning to the null surface point of view, we claim
that any solutions $\nu=\nu(\theta^i,\z,\bar\z)$ to Eq.(\ref{e}) {\it
and} the metricity conditions as well (since the metricity conditions
guarantee a metric), will provide us with a Lorentzian metric that
satisfies the \underline{complete} set of Einstein equations. It is in
this sense that we say that Eq.(\ref{e}) encodes the full Einstein
equations.

[We mention that the only piece of information that we lack in the null
surface approach is the integration constant of the Bianchi
identities.  This integration constant is the cosmological constant in
the vacuum case, or part of the stress tensor in the case of
non-vanishing sources.]

In the remainder of this section we will clarify the issues just
raised. First, we will return to the conventional form of the Einstein
equations and calculate the trace of the Einstein Eqs.(\ref{EM}). Then
we will show that in the null surface approach, this trace equation is
satisfied if Eq.(\ref{e}) and the metricity conditions are satisfied.
Thus, we will have proved our claim that we do not need the trace
equation. In the null surface approach, the tenth equation is thus
encoded into the metricity conditions and (\ref{e}).

Contracting Eq.(\ref{EM}) with $\eta^{ab}$ we obtain:  \be
	-3\Box\nu + \frac12 \p^i\p^jf_{ij} = T.         \label{tracef}
\ee By virtue of our previous calculation for the double divergence of
the metric $f_{ij}$, i.e. Eq.(\ref{doublediv}), Eq.(\ref{tracef}) turns
out to be an equation for only the conformal factor $\nu$, decoupled
from $\Ld$:

\be -3\Box\nu - \p_1^2\bar\eth\eth\nu = T           \label{traceq} \ee
where $T$ is uniquely given by $\p_aT = -4\p^bt_{ab}$ up to a
constant.  This would be the tenth Einstein equation. We now show that
(\ref{traceq}) is satisfied by virtue of (\ref{e}). With some work (by
essentially using the commutation relations (\ref{cr})) it can be shown
that Eq.(\ref{e}) implies the following:

\be \p_1(-3\Box\nu - \p_1^2\bar\eth\eth\nu) = \p_1T \ee (where $\p_1T$
is equal to $-4\p^jt_{1j})$. Furthermore, we show in the appendix that
the metricity conditions (*) and (A) imply:  \be \eth(-3\Box\nu -
\p_1^2\bar\eth\eth\nu) = 0     \label{ethtrace} \ee \be
\bar\eth(-3\Box\nu - \p_1^2\bar\eth\eth\nu) = 0. \label{barethtrace}
\ee These are sufficient conditions to guarantee that $\p_i(-3\Box\nu -
\p_1^2\bar\eth\eth\nu) = \p_iT$ for all $i$ because the application of
$\eth$ and $\bar\eth$ on $\p_1(-3\Box\nu-\p_1^2\bar\eth\eth\nu) =
\p_1T$ yields the remaining equations if $(-3\Box\nu -
\p_1^2\bar\eth\eth\nu)$ is a function only of $x^a$, as implied by
(\ref{ethtrace}) and (\ref{barethtrace}).  Therefore any solution to
Eq.(\ref{e}) and the metricity conditions satisfies the tenth equation
automatically. Note that (\ref{ethtrace}) and (\ref{barethtrace}) imply
severe conditions on $\alpha$ and $\beta$ in the solution to (\ref{e})
(see Sec.IV.D).

\subsection{The solution to the metricity conditions}

Our point of view is now as follows: without yet worrying about the
details (see discussion at the end of this Sec.) the conformal factor
$\nu$ can be found explicitly as a double integral in $R$ of the given
stress tensor $t_{11}$. Once $\nu$ has been solved for, it enters the
equations for $\Ld$ (Eq.(\ref{ethg01}), (\ref{A}) and (\ref{C})) as a
known source only. Our task is then to solve the three kinematic
equations for $\Ld$. [It should be noted, however, that the decoupling
of the equations for $\nu$ and $\Ld$ can not be carried out in this way
in the full theory, in which the metricity conditions and the dynamical
equation are strongly coupled~\cite{fkn95}.]

We begin by manipulating (*) and (A) to obtain two very important
relations for $\Ld$. The following is merely the derivation of
Eqs.(\ref{Ted's}) and (\ref{box}) which we include at this point for
the sake of completeness (too frequently in this work the results are
not easy to reproduce, due in part to the non trivial commutation
relations (\ref{cr})).

By applying $\bar\eth$ to (\ref{A}) and commuting $\bar\eth \eth$ on
the right side we obtain:

\[ \bar\eth\Ld_- = \frac12\eth\bar\eth\Ld_1 + 2 \Ld_1 \]

By applying now $\p_1$ to both sides and commuting $\p_1\bar\eth$ in
the left side we obtain:  \[ \bar\eth\p_1\Ld_- + \p_+\Ld_- = \frac12
\p_1\eth\bar\eth\Ld_1 +
					2\p_1\Ld_1.
\] If we commute now $\bar\eth\p_-$ in the left side we obtain:  \[
\p_-\bar\eth\Ld_1 - \p_0\Ld_1 + 2\p_1\Ld_1 + \p_+\Ld_- =
			\frac12 \p_1\eth\bar\eth\Ld_1 + 2\p_1\Ld_1.
\]

The $\p_1\Ld_1$ cancel out from both sides. We now use the basic
equation (\ref{ethg01}) to eliminate $\bar\eth\Ld_1$ in the left side
and obtain:

\be \label{+-} \p_+\Ld_- = -\p_1\Ld_0 - \frac12 \p_1\eth\bar\eth\Ld_1 +
4\p_-\eth\nu.  \ee

Next we carry out a similar procedure on (\ref{ethg01}). First we apply
$\eth$ and then $\p_1$ to both sides.  By commuting $\p_1\eth$ on the
left side obtain:  \[ \eth\p_1\Ld_+ + \p_-\Ld_+ = - \frac12
\p_1\eth\bar\eth\Ld_1
				+ 2\p_1\eth^2\nu.
\] Further, by commuting $\eth\p_+$ on the left side we obtain:  \[
\p_+\eth\Ld_1 - \p_0\Ld_1 + 2\p_1\Ld_1 + \p_-\Ld_+ = - \frac12
\p_1\eth\bar\eth\Ld_1 + 2\p_1\eth^2\nu.  \] Using (\ref{A}) to
eliminate $\eth\Ld_-$ on the left side we are left with:  \[
 3\p_-\Ld_+  - \p_0\Ld_1 + 2\p_1\Ld_1= - \frac12 \p_1\eth\bar\eth\Ld_1
			+ 2\p_1\eth^2\nu.
\] Using $\p_+\Ld_-=\p_-\Ld_+$ and inserting $\p_+\Ld_-$ from
(\ref{+-}), we obtain our \underline{first basic result}:

\be 4\p_1\Ld_0 + \p_1\eth\bar\eth\Ld_1 - 2\p_1\Ld_1 = 12\p_-\eth\nu -
2\p_1\eth^2\nu                          \label{Ted's} \ee

Eliminating $\p_1\eth\bar\eth\Ld_1$ from (\ref{Ted's}) via (\ref{+-})
we obtain our \underline{second basic result}:

\be \Box\Ld \equiv 2\bigl(\p_0\p_1-\p_1\p_1-\p_+\p_-\bigr)\Ld
	= 4\p_-\eth\nu -2\p_1\eth^2\nu.
	\label{box} \ee $\Ld$ thus satisfies the wave equation with the
conformal factor playing the role of the source. [This equation,
(\ref{box}), has been derived here just for completeness and will not
be used. Its use will be presented in a future paper where $\Ld$ will
be the basic variable and $Z$ purely auxiliary. In the present paper we
use $\Ld$ as auxiliary while $Z$ plays the primary role.]

Equation (\ref{Ted's}) can be integrated on $R$, assuming
asymptotically flat boundary conditions, to yield the following:

\be \Ld_0 = -\frac14\eth\bar\eth\Ld_1 +\frac12\Ld_1 +
3\int^{R}_{\infty}\!\p_-\eth\nu d\tilde R - \frac12\eth^2\nu +
\sigma,_0(u,\z,\bar\z)                          \label{Ted's0} \ee
where $\sigma(u,\z,\bar\z)$ is free radiation data.

Taking $\bar\eth$ of $\Ld_0$ and inserting it into the third metricity
condition (\ref{C}) we find:

\be -\frac14\eth^3\bar\Ld_1 = \bar\eth\sigma,_0 +
3\bar\eth\!\int^{R}_{\infty}\!\p_-\eth\nu d\tilde R -2\eth\nu
-\frac32\eth^2\bar\eth\nu                           \label{eth3} \ee
and its complex conjugate.

By applying $\bar\eth$ to (\ref{C}) we obtain:  \[ \bar\eth^2\Ld_0 =
-\frac14\bar\eth\eth^3\bar\Ld_1 -\frac14\eth\bar\eth^3\Ld_1
+\bar\eth\eth\nu +\eth\bar\eth\eth\bar\eth\nu.  \] Now using
(\ref{eth3}) and its complex conjugate in the previous equation leads
to:

\be \bar\eth^2\Ld_0 = \bar\eth^2\sigma,_0+\eth^2\bar\sigma,_0
+3\bar\eth^2\!\int^{R}_{\infty}\!\p_-\eth\nu d\tilde R
+3\eth^2\!\int^{R}_{\infty}\!\p_+\bar\eth\nu d\tilde R-3\bar\eth\eth\nu
-2\bar\eth\eth^2\bar\eth\nu \ee

Integrating once in $u$ and using $\Ld=\eth^2Z$ finally yields the
\underline{fundamental equation} for the cut function $Z$:

\be \bar\eth^2\eth^2Z = \bar\eth^2\sigma + \eth^2\bar\sigma
+\int^{u}\!\biggl(3\bar\eth^2\!\int^{R}_{\infty}\!\p_-\eth\nu d\tilde R
+3\eth^2\!\int^{R}_{\infty}\!\p_+\bar\eth\nu d\tilde R
-3\bar\eth\eth\nu -2\bar\eth\eth^2\bar\eth\nu\biggr)d\tilde
u.           \label{z} \ee

Equation (\ref{z}), our final equation, equivalent to the linearized
Einstein equations, can be solved by means of a simple Green's function
for the operator $\bar\eth^2\eth^2$ (essentially the double laplacian
on the sphere). The Green's function is given explicitly in appendix
C~\cite{ikn89}. As complicated as it appears, the right hand side is
simply the source, a known function of $(x^a,\z,\bar\z)$ once the
conformal factor has been solved for.

To solve for the conformal factor in terms of only the source $t_{11}$
is a task more subtle than it appears at first sight. Formally, one can
write the solution to Eq.(\ref{e}) as the double integral of $t_{11}$
plus a homogeneous solution that is linear in $R$ and that contains the
values of $\nu$ and $\p_1\nu$ at a fixed value $R=R_0$, as shown
earlier. The homogeneous part is severely restricted, however, by the
trace equation. The trace equation, in terms of $\alpha$ and $\beta$
becomes quite cumbersome in the case of general (arbitrary) sources,
involving the values of $t_{11}$ and different derivatives of $t_{11}$
in a non-trivial manner at the surface $R=R_0$. We will bypass the
problem of solving for the homogeneous part of the conformal factor in
the general case by restricting ourselves to the case where {\it there
exists a surface $R=R_0$ on which $t_{11}$ vanishes as a function of
$(u,R_0,\omega,\bar\omega,\z,\bar\z)$}. This includes the
asymptotically flat case (where $t_{11}=0$ at $\cal J^+$) but is
slightly more general.

For the particular case of sources that vanish at a surface $R=R_0$,
the conformal factor can be solved for unambiguously by:  \be
\nu=\int^{R}_{R_0}\int^{\tilde R}_{R_0}\!t_{11}dR'd\tilde R  \label{nu}
\ee in the gauge such that $\alpha=\beta=0$ (see next section for a
discussion of gauge).

The corresponding equation for the cut function is the same as
(\ref{z}) with the lower limit of the integrals being replaced by
$R_0$.

\subsection{Gauge conditions and remaining freedom}

This section is supplementary to the linearized null surface approach
described in the previous sections. We justify particular choices of
the conformal factor $\nu$ made in past works~\cite{i93} (and in
(\ref{nu})) and provide some insights into the issue of the gauge
freedom.

We mentioned before, Sec.II, that by imposing $q_{1b}=0$ we were still
left with gauge freedom of the form:  \[ \p_1\xi_i + \p_b\xi_1 -
\frac12\eta_{1i}\p^k\xi_k = 0.  \] For $i=1$ we have:  \[ \xi_1,_1 = 0
\] therefore $\xi_1=\phi(u,\omega,\bar\omega,\z,\bar\z)$. For $i=+$ we
have:  \[ \xi_1,_+ + \; \xi_+,_1 = 0 \] therefore $\xi_+ = -\phi,_+R +
\tau_+$ where $\tau_+$ does not depend on $R$.  Similarly for $i=-$ we
obtain $\xi_- = -\phi,_-R + \tau_-$ where $\tau_-$ does not depend on
$R$. The component $\xi_0$ can be solved for in a similar fashion from
$\p_1\xi_0 + \p_0\xi_1 - \frac12\p^k\xi_k = 0$ but it will not be
needed in what follows. The divergence $\p^k\xi_k$ is equal to
$-2(\xi_+,_- + \xi_-,_+)$ and therefore $\p^k\xi_k=4\phi,_{+-}R
-2(\tau_+,_- + \tau_-,_+)$. At this stage it is clear that the gauge
freedom in $\nu$ (see (\ref{gaugenu})) is linear in $R$ with two free
functions of the remaining coordinates (and $(\z,\bar\z)$) as the
coefficients.

Additional gauge conditions arise from our $(\z,\bar\z)$-dependent
coordinate systems; see Eqs.(\ref{coord}). The field $\xi^i$ is just an
infinitesimal variation of the coordinates $\theta^i$ to neighboring
coordinates $\theta'^i=\theta^i + \xi^i$. Explicitly:  \bea u'& = &
u+\delta u = u+\xi^0                            \nonumber\\ \omega' & =
& \omega+\delta\omega = \omega+\xi^+        \nonumber\\ \bar\omega' & =
& \bar\omega +\delta\bar\omega =
			\bar\omega+\xi^-                \nonumber\\
R' & = & R  + \delta R = R + \xi^1.  \eea Since $\theta'^i$ and
$\theta^i$ for $i=1,+,-$ are obtained from $\theta'^0$ and $\theta^0$
by the application of $\eth$, $\bar\eth$ and $\eth\bar\eth$, the
components $\xi^+$, $\xi^-$ and $\xi^1$ can also be obtained by
applying $\eth$ and $\bar\eth$ to $\xi^0$, i.e.:  \bea \xi^+ & =
&\eth\xi^0                            \nonumber\\ \xi^- & =
&\bar\eth\xi^0                        \nonumber\\ \xi^1 & = &
\bar\eth\eth\xi^0.                  \nonumber \eea Thus, $\tau_+$,
$\tau_-$ and $\phi$ are not independent of each other. If we lower the
index with $\xi_i=\eta_{ij}\xi^j$ we find the following:

\[ \xi_1=\xi^0=\phi(u,\omega,\bar\omega,\z,\bar\z) \] \[ \xi_+=-\xi^- =
-\bar\eth\xi^0 =  -\bar\eth\xi_1 \] \[ \xi_-=-\xi^+ = -\eth\xi^0 =
-\eth\xi_1.  \] Since $\xi_1=\phi(u,\omega,\bar\omega,\z,\bar\z)$ and
the coordinates $u$, $\omega$ and $\bar\omega$ depend on $(\z,\bar\z)$
then taking $\eth$ corresponds to taking a total derivative with
respect to $(\z,\bar\z)$. Assuming that $\phi$ is a first order
quantity we have:  \bea \eth\xi_1 = \eth\phi & = & \phi,_0\eth \theta^0
+ \phi,_+\eth\theta^+ + \phi,_-\eth\theta^-+ \eth'\phi  \nonumber\\
		     & = & \phi,_0\omega + \phi,_-R + \eth'\phi
							\nonumber\\
\bar\eth\xi_1=\bar\eth\phi & = & \phi,_0\bar\eth\theta^0 +
\phi,{+-}\bar\eth\theta^+ + \phi,_-\bar\eth\theta^- +
\bar\eth'\phi                 \nonumber\\
			   & = & \phi,_0\bar\omega +\phi,_+R
			   +\bar\eth'\phi
							\nonumber
\eea where $\eth'$ and $\bar\eth'$ are taken \underline{holding
$\theta^i$ fixed} and the definitions of $\omega$, $\bar\omega$ and $R$
have been used.
  Therefore

\[ \tau_+ = -\phi,_0\bar\omega - \bar\eth'\phi \] \[ \tau_- =
-\phi,_0\omega - \eth'\phi.  \] With these expressions the divergence
$\p^k\xi_k$ is given as:  \be \p^k\xi_k = 4\phi,_{+-}R + 4\phi,_0 +
2\phi,_{0-}\bar\omega
	    +2\phi,_{0+}\omega +2\bar\eth'\phi,_-
				    +2\eth'\phi,_+.   \label{gauge}
\ee Here $\phi$, an arbitrary function of
$(u,\omega,\bar\omega,\z,\bar\z)$, represents our full gauge freedom.
The divergence $\p^k\xi_k$ can be used to modify $\nu$, as noted in
Sec.II.  Recall (\ref{gaugenu}):  \[
 \nu' = \nu - \frac14 \p^a\xi_a.  \]

By virtue of Eq.(\ref{e}), $\nu$ is determined by $t_{11}$ up to a
linear function of $R$. We are interested in seeing whether we can
gauge away the linear (homogeneous) part of $\nu$. We claim that if
$\nu$ satisfies Eq.(\ref{e}) and the metricity conditions, then we have
just enough freedom to gauge away the homogeneous part of $\nu$, in the
case where the cosmological constant vanishes and there is a surface
$R=R_0$ on which $t_{11}$ vanishes. In particular, {\it $\nu$ would be
pure gauge} for the vacuum case with vanishing cosmological constant.
This would justify the choice $\nu=0$ made in earlier works on the
linearized vacuum null surface theory~\cite{i93}.

The proof of our claim is in the following argument. As was previously
shown, any $\nu$ that satisfies both Eq.(\ref{e}) and the metricity
conditions will consequently satisfy \[ -3\Box\nu
-\p_1^2\eth\bar\eth\nu = T, \] the trace equation. Although not needed
in general, it proves particularly useful in this section. If there
exists a surface $R=R_0$ in which $t_{11}$ vanishes, when we substitute
$\nu=\int^{R}_{R_0}\int^{\tilde
R}_{R_0}\!t_{11}dR'd\tilde{R}+\alpha+\beta R$ (the general solution to
Eq.(\ref{e})) into the trace equation, the non-homogeneous part of
$\nu$ drops out and we obtain an equation for only $\alpha$ and $\beta$
that has the following form:  \be 8\dot{\beta} - 4\alpha,_{+-} +
2\dot{\beta},_-\bar\omega + 2\dot{\beta},_+\omega + 2\bar\eth '\beta,_-
+ 2\eth '\beta,_+ =
					- \ld           \label{traceab}
\ee where $\eth'$ and $\bar\eth'$ are taken by holding the coordinates
$\theta^i$ fixed, $\dot{\beta}\equiv\beta,_0$ and $\ld$ is the
cosmological constant.

One technical point that was needed in the derivation of
(\ref{traceab}) is the commutator of $\eth$ through the integral sign,
given by (boundary terms vanish for our particular restriction on
$t_{11}$):

\[ \eth\int^{R}_{R_0}\!\!F d\tilde R=\int^{R}_{R_0}\!\!\eth F d\tilde
R+ \int^{R}_{R_0}\!\!\!\int^{\tilde R}_{R_0}\!\!\p_-F dR'.  \]

In this sense, we think of the trace equation as an equation for just
the homogeneous part of $\nu$.  It is remarkable that $\p^k\xi_k$
(given above (\ref{gauge}) of the form $\tilde\alpha + \tilde\beta R$)
is of the form so that $\tilde\alpha$ and $\tilde\beta$ {solve} the
trace equation. In other words, the expression for $\p^k\xi_k$ can be
chosen as the solution to the trace equation (with vanishing
cosmological constant).  Therefore, we can always choose a gauge so
that the conformal factor $\nu$ has no homogenous part, if the
cosmological constant is zero and the source $t_{11}$ vanishes on a
surface $R=R_0$.

The solution to the trace equation with non-vanishing $\ld$, modulo
gauge, is

\be \nu = \ld (uR - R^2 -\omega\bar\omega).  \ee This solution can not
be gauged away. We are free to choose $\ld$ to vanish or not to vanish.
A general way to rule out a possible cosmological constant is by
requiring asymptotically flat boundary conditions for the conformal
factor.

\section{concluding remarks}

The null surface theory approach to GR to first order just described is
meant to serve as a reliable test of the concepts of the full null
surface theory. We found it extremely useful in clarifying the full
version of the theory, which will be published elsewhere~\cite{fkn95}.
As a restriction to the linearized regime, however, it's utility is
limited and some results can not be carried over to the full theory.
For example, we found that condition (B) in Sec.(IV.A.) is an identity
in first order, but the fact that it remains an identity to all orders
can not be established from the linearization. Nevertheless, it gave a
hint as to what to expect from the full theory with respect to the
relevance of (B).

The linearized null surface theory departs conceptually from the full
theory in that it is a theory for the quantity $\Ld$ {\it on a fixed
known background}. In this sense, many of the complications of the full
theory do not have a place here, in particular, the difficulty that the
coordinates and $\Ld$ are conceptually intertwined in the full theory,
which makes the equations extremely lengthy (and the ideas at the start
quite confusing). It's a fact that the linearization of the null
surface theory becomes itself a separate theory, easier to handle,
where the methods used do not need to agree completely with the ones
needed in the full theory. For example, the conformal factor $\nu$ can
actually be solved for without knowing $\Ld$, and then it can be used
as a known source to find $\Ld$, a fact that does not apply to the full
theory.

Throughout this work, the null surfaces are known to zeroth order and
get corrected by the first order $\Ld$, i.e., $\Ld$ gives a first order
correction to the null planes introduced in Sec.II. One can develop a
perturbative scheme in which the surfaces are null planes in flat
space, and get corrected successively with every higher order step in
the hierarchy. The complications in the equations increase considerably
at the second order level, making it, at least for the moment, a very
difficult task to accomplish explicitly . A formal procedure can be
outlined which shows no particular complications in the development of
a perturbative scheme around the flat null planes.

First order vacuum theory can be obtained by setting the sources equal
to zero, and taking $\nu=0$. It is interesting to notice that under
these conditions, $f_{ij}$ satisfies both the following:
$\eta^{ij}f_{ij}=0$ and $\p^jf_{ij}=0$, from equations (\ref{div0}) and
(\ref{div-}). This means that in vacuum our gauge conditions are a null
version of the harmonic gauge. Furthermore, equation (\ref{box})
reduces to the homogenous wave equation $\Box\Ld=0$. Since all the
components of $f_{ij}$ can be found by applying partial derivatives and
$\eth$ or $\bar\eth$ operations to $\Ld$ (all of which commute with
$\Box$) we have the result that every component of the metric satisfies
the homogeneous wave equation, $\Box f_{ij}=0$. We are currently
exploring the possibility of finding the general solution to (*), (A)
and (C) in vacuum directly for $\Ld$ without the use of $Z$. This
approach, which reverses the roles of $Z$ and $\Ld$ as primary versus
auxiliary quantities, and which yields new insights into the metricity
conditions, will be presented elsewhere.

\acknowledgements We gratefully thank Savi Iyer, Al Janis and Lionel
Mason for enlightening discussion. Support from CONICET and from the
NSF under grants \# PHY 89-04035 and PHY 92-05109 is also gratefully
acknowledged.

\appendix

\section{The solution of the cut function equation by means of Green's
functions}

The equation for the linearized cut function (\ref{z}) is of the form:
\be \bar\eth^2\eth^2Z = S(x^a,\z,\bar\z) \ee where $Z$ is a function of
$(\z,\bar\z)$ and the four parameters $x^a$.  $S$ is a given source.
The solution to this equation is of the form:  \be Z(\z,\bar\z,x^a) =
\int G(\z,\bar\z,\z',\bar\z')S(\z',\bar\z') dS' \ee where
$G(\z,\bar\z,\z',\bar\z')$ is the appropriate Green's function for the
operator $\bar\eth^2\eth^2$ on the sphere. This function, which has
been found~\cite{i93}, is a rather simple expression in terms of the
vector $l^a$ introduced earlier in this work (see (\ref{l}):  \be
G(\z,\bar\z,\z',\bar\z') =
\frac{1}{4\pi}l^a(\z,\bar\z)l_a(\z',\bar\z')ln\Bigl(l^b(\z,\bar\z)l_b(\z',\bar\z')\Bigr).
\ee It is also useful to have the explicit version in terms of
$(\z,\bar\z)$, which is found from:  \[ l^a(\z,\bar\z)l_a(\z',\bar\z')
= \frac{(\z-\z')(\bar\z-\bar\z')}{(1+\z\bar\z)(1+\z'\bar\z')}.  \]

There is a method for explicitly evaluating integrals of this type on
the sphere for a large class of sources $S$~\cite{knp84}. This method
consists of rewriting the integral as a sum of contour integrals around
simple poles. In this way the integral is reduced in a large part to
the calculation of residues. The method works for singular integrands
as well, our present case. Explicit results have already been obtained
by the authors for the vacuum case with arbitrary quadrupole radiation
data of the form $\sigma(u,\z,\bar\z) = a(u) _2Y_{2m}(\z,\bar\z)$,
where $ _2Y_{2m}(\z,\bar\z)$ is a spin-weight-2 spherical harmonic and
$a(u)$ is {\it arbitrary}. These results will appear elsewhere.

\section{calculation of the divergence and double divergence of the
metric}

For the component $i=-$ only the use of (\ref{ethg01}) is needed:  \bea
\p^jf_{-j} & = & \p^0f_{-0} + \p^-f_{--}
\nonumber\\
	   & = & \p^0(\frac12\bar\eth\Ld_1)+\p^-(-\Ld_1)
	   \nonumber\\ & = &
	   \p_1(\frac12\bar\eth\Ld_1)-\p_+(-\Ld_1)      \nonumber\\ & =
	   & \p_1\bigl(\frac12\bar\eth\Ld_1 + \Ld_+\bigr) \nonumber\\ &
	   = & 2\p_1\eth\nu .
\eea

The component $i=0$ requires the use of (\ref{ethg01}) and the
commutation relations:

\bea \p^jf_{0j} & = & \p^0f_{00} + \p^-f_{0-} + \p^+f_{0+}
\nonumber\\
	   & = & \p^0(-\frac12\bar\eth^2\Ld_1) + \p^-f_{0-}
		 + \p_1\eth\bar\eth\nu + complex\;\;conjugate.
		 \nonumber\\ & = & \p^0(-\frac12\bar\eth^2\Ld_1) +
	   \p^-(\frac12\bar\eth\Ld_1)
		 + \p_1\eth\bar\eth\nu+ complex\;\;conjugate.
		 \nonumber\\ & = & (-\p^0\bar\eth + \p^-)
	   \frac12\bar\eth\Ld_1
		 + \p_1\eth\bar\eth\nu+ complex\;\;conjugate.
		 \nonumber\\ & = & -(\p_1\bar\eth + \p_+)
	   \frac12\bar\eth\Ld_1
		 + \p_1\eth\bar\eth\nu + complex\;\;conjugate.
		 \nonumber\\ & = & -(\p_1\bar\eth +
	   \p_+)(2\eth\nu-\Ld_+)
		 + \p_1\eth\bar\eth\nu + complex\;\;conjugate.
		 \nonumber\\ & = & -2(\p_1\bar\eth +\p_+)\eth\nu +
		  \p_+(\p_1\bar\eth + \p_+)\Ld + \p_1\eth\bar\eth\nu +
		 complex\;\;conjugate.  \nonumber\\ & = &
	   -2(\p_1\bar\eth + \p_+) \eth\nu +
		  \p_+(\bar\eth\p_1 + 2\p_+)\Ld + \p_1\eth\bar\eth\nu +
		 complex\;\;conjugate.  \nonumber\\ & = &
	   -2(\p_1\bar\eth + \p_+) \eth\nu +
		  \p_+(4\eth\nu) + \p_1\eth\bar\eth\nu +
		 complex\;\;conjugate.  \nonumber\\ & = &
	   -2\p_1\bar\eth\eth\nu + 2\p_+\eth\nu
		 + \p_1\eth\bar\eth\nu + complex\;\;conjugate.
		 \nonumber\\ & = & -2\p_1\bar\eth\eth\nu + 2\p_+\eth\nu
					+2\p_-\bar\eth\nu .
\eea

The double divergence $\p^i\p^jf_{ij}$ becomes:  \bea
 \p^i\p^jf_{ij} & = & \p^0\p^jf_{0j} + \p^+\p^jf_{+j}
				    + \p^-\p^jf_{-j}
				    \nonumber\\
		& = & \p_1\p^jf_{0j} - \p_-\p^jf_{+j}
				    - \p_+\p^jf_{-j}
				    \nonumber\\
		& = & \p_1(-2\p_1\bar\eth\eth\nu + 2\p_+\eth\nu
			+2\p_-\bar\eth\nu) - 2\p_-(\p_1\bar\eth\nu) -
			2\p_+(\p_1\eth\nu)
			\nonumber\\ & = & -2\p_1^2\bar\eth\eth\nu.
\eea

\section{The stress energy tensor from the null surface point of view}

Since we have an explicit $(\z,\bar\z)$ dependence of the component
$t_{11}$ of a given symmetric trace-free tensor $t_{ij}$, the remaining
components can be obtained by an appropriate combination of successive
applications of $\eth$ and $\bar\eth$ to $t_{11}$.  The following have
been obtained with the help of the vanishing trace condition, $t_{11} -
t_{01} + t_{+-} = 0$:  \[ t_{1-} = -\frac12\eth t_{11} \] \[ t_{1+} =
-\frac12\bar\eth t_{11} \] \[ t_{--} = \frac12\eth^2 t_{11} \] \[
t_{++} = \frac12\bar\eth^2 t_{11} \] \[ t_{+-} = \frac14\eth\bar\eth
t_{11} + \frac12 t_{11} \] \[ t_{0-} = -\frac14\bar\eth \eth^2 t_{11} -
\eth t_{11} \] \[ t_{0+} = -\frac14\eth\bar\eth^2 t_{11} - \bar\eth
t_{11} \] \[ t_{00} = \frac14\bar\eth^2\eth^2 t_{11} +
\frac32\eth\bar\eth t_{11} + 3t_{11} \] \be t_{01} = \frac14
\eth\bar\eth t_{11} + \frac32 t_{11}   \label{tij} \ee

As an example of the calculation, the first of (\ref{tij}) was obtained
in the following manner:
$t_{11}=\theta_1^a\theta_1^bt_{ab}=l^al^bt_{ab}$ therefore $\eth
t_{11}=2\eth
l^al^bt_{ab}=2m^al^bt_{ab}=-2\theta_-^a\theta_1^bt_{ab}=-2t_{1-}$ and
inverting: $t_{1-}=-\frac12\eth t_{11}$.

Since the components of the trace-free part of the stress tensor
constitute the right side of the trace-free Einstein equations, then by
the same operations suggested above for $t_{ij}$ we can obtain the nine
trace-free Einstein equations from $\p_1^2\nu = t_{11}$.

\section{the trace of the Einstein system and the metricity conditions}

Here we show that \[ \eth\Bigl(3\Box\nu + \p_1^2\eth\bar\eth\nu \Bigr)
= 0 \] by virtue of the metricity conditions. Though the calculation is
fairly lengthy, we will outline the argument. We need Eq.(*), (A) and
Eq.(\ref{box}):  \be \Box\Ld = 4\p_-\eth\nu -2\p_1\eth^2\nu =
-2(\p_1\eth-2\p_-)\eth\nu
							\label{waveld}
\ee (obtained from (*) and (A)).

First we turn $\eth\p_1^2\eth\bar\eth\nu $ into a more suitable
expression:

\[ \eth\p_1^2\eth\bar\eth\nu = \p_1\bar\eth(\p_1\eth-2\p_-)\eth\nu +
\p_1(\p_+\eth+2\p_1-2\p_0)\eth\nu.  \] Now we use (\ref{waveld}) to put
the first term in terms of $\Box\Ld$ and use (*) to eliminate $\nu$ in
favor of $\Ld$ (also commute $\p_1\bar\eth$ in the first term):  \[
\eth\p_1^2\eth\bar\eth\nu = -\frac12\Box\bar\eth\Ld_1 -
\frac12\Box\Ld_+ + \frac12\p_1(\p_+\eth+2\p_1-2\p_0)(\Ld_+ +
\frac12\bar\eth\Ld_1).  \] If we put it together and after some work we
obtain:  \[ \eth\Bigl(3\Box\nu +\p_1^2\eth\bar\eth\nu \Bigr) =
\frac12\p_+\Bigl(\Box\Ld + (\p_1\eth-2\p_-)(\Ld_+ +
\frac12\bar\eth\Ld_1)\Bigr).  \] If we commute $\bar\eth$ and $\p_-$ in
the term $\p_-\bar\eth\Ld_1$ and use (*) to eliminate $\Ld_-$ in favor
of $\eth\Ld_1$ we obtain:

\bea
 \eth\Bigl(3\Box\nu +\p_1^2\eth\bar\eth\nu \Bigr) & = &
\frac12\p_+\Bigl(\Box\Ld + (\p_1\eth-2\p_-)\Ld_+ +
\frac12\p_1\eth\bar\eth\Ld_1)                   \nonumber\\
	&   & -\frac12\p_1\bar\eth\eth\Ld_1 + \frac12\p_+\eth\Ld_1
		- \p_0\Ld_1 +2\p_1\Ld_1\Bigr).  \nonumber
\eea With some more work using the commutation relations only we
obtain:  \[
 \eth\Bigl(3\Box\nu +\p_1^2\eth\bar\eth\nu \Bigr) = \frac12\p_+\Bigl(
			 -3\p_+(\Ld_--\frac12\eth\Ld_1) +
		\frac12\p_1(\eth\bar\eth -\bar\eth\eth +4)\Ld_1
		\Bigr).
\] This last result is clearly vanishing by virtue of the metricity
condition (A) and the commutation relations for $\eth$ and $\bar\eth$
on quantities of spin weight equal to 2.

\end{document}